\documentstyle{article}
\begin{document}
\LARGE
\begin{center}
\bf  Entropy of Constrained Gravitational Instanton
\vspace*{0.6in}
\normalsize \large \rm

Wu Zhong Chao

Dept. of Physics

Beijing Normal University

Beijing 100875, China

(Dec. 15, 1998)

\vspace*{0.4in}
\large
\bf
Abstract
\end{center}
\vspace*{.1in}
\rm
\normalsize
\vspace*{0.1in}

The seeds for quantum creations of universes are constrained
gravitational instantons. For all compact constrained instantons
with a $U(1)$
isometry,  the period $\beta$ of the
group parameter $\tau$ is identified as the reciprocal of the
temperature. If $\beta$ remains a free parameter under the
constraints, then  the Euclidean action
becomes the negative of the ``entropy.'' As examples, we perform
the calculations for the Taub-NUT
and Taub-Bolt-type models and study the quantum creation of the
Taub-NUT
universe.

\vspace*{0.3in}

PACS number(s): 98.80.Hw, 98.80.Bp, 04.60.Kz, 04.70.Dy


\vspace*{0.3in}

e-mail: wu@axp3g9.icra.it

\pagebreak

\vspace*{0.3in}

In the No-Boundary Universe [1], it was previously thought that
the seed for the creation of the
universe must be an instanton. Recently, it was realized that
this argument can only be applies to the
creations with stationary creation probabilities. For a general
creation, at the $WKB$ level, the
seed for a created universe must be a constrained gravitational
instanton [2][3].

The constrained gravitational instanton approach has been
successfully used in treating
the problem of quantum creations of the Kerr-Newman black hole
pairs in the (anti-)de Sitter 
space background [4]. For the chargeless and nonrotating case,
the creation probability is the
exponential of the (minus) entropy of the universe. For the other
cases (charged, rotating, or both),
the creation probability is the exponential of (minus) one
quarter of the sum of the inner and outer
black hole horizon areas. The case of the Kerr-Newman family is
also solved as a limiting case of
that for the Kerr-Newman-anti-de Sitter family. It is noted that
the problem of the open creation is
solved in the absence of a general no-boundary proposal for open
creation.

For the pair creation of black holes, the constrained instanton
is constructed through cutting,
folding and pasting of two imaginary time $\tau$ sections passing
two horizons in the complex black hole solution. The coordinate
$\tau$ is also the group parameter
of the $U(1)$ isometry. Its identification period $\beta$ is the
reciprocal of the temperature. The manifold satisfies the
Einstein
and other field equations everywhere except for the horizons. 
The parameter $\beta$ is the
only degree of freedom left.  It is surprising that the action
$I$ of the instanton is independent of
$\beta$. Therefore, the manifold is a stationary action orbit and
can be used for the $WKB$
approximation for the wave function and probability. The relative
probability is the exponential of
the negative of the action of the constrained instanton, as 
described above. The Lorentzian spacetime is obtained through the
analytic continuation of the coordinate $\tau$ from a real value
to an imaginary value.

In Euclidean quantum gravity, the background entropy $S$ can be
evaluated from the action [5]
\begin{equation}
S = - \frac{\beta \partial}{\partial \beta} \ln Z + \ln Z .
\end{equation}
The independence of our action from $\beta$ implies that the
entropy is the negative of the action.

In fact, this argument is also valid for the general case
of $U(1)$ isometric constrained instantons, when
constructed as follows. We begin with a complex manifold 
satisfying the
Einstein equation everywhere
except for some singularities. The group parameter $\tau$ is
identified as an imaginary time, as in the above black hole
case. We assume that its period $\beta$ is one of the free
parameters
characterizing the singularities under the constraints for the
instanton. For the manifold to be
qualified as a constrained instanton, its action must be
stationary with respect to all these free
parameters left. Thus, eq. (1) demonstrates that the stationary
property of the action with
respect to $\beta$ implies that the entropy is the negative of
the action. However, one has to clarify the meaning of the
entropy here [4].

It is interesting to note that if there are at least two $U(1)$
isometries and their associated,
different temperatures as free parameters under the same
constraints, like the Taub-NUT and
Taub-Bolt-type models discussed below, then the entropy
is unique and independent of the choice of  temperatures. 

When one is studying a system with constrained quantities,
one has to use the grand
partition function $Z$ in grand canonical ensembles for
gravitational thermodynamics. For
example, if the system is constrained by three quantities, namely
mass or energy $m$, electric
charge $Q$ and angular momentum $J$, then the partition function
over the metric $g$ and matter fields $\chi$ is [5]
\begin{equation}
Z = \mbox{Tr} \exp(-\beta m + \beta \Omega J + \beta \Phi Q) =
\int d [g] d [\chi ] \exp-I(g, \chi ),
\end{equation}
where $\Phi$ and $\Omega$ are the corresponding chemical
potentials and the path integral is
over all fields with the junction condition that their value at
the
point $(\tau - \beta, \phi + i \beta
\Omega)$ is $\exp(Q\beta \phi)$ times the value at $(\tau,
\phi)$, where $\phi$ is the coordinate of
the rotation. At the $WKB$ level, the path integral is the
exponential to the negative of the action
of the constrained instanton. We shall consider the case with a
compact instanton. Because the
total energy vanishes, one can set $m$ to be zero.

However, there does not exist a constrained instanton under the
junction condition for the pasting with nonzero $Q$
or $J$. It turns out that the $WKB$ approximation of the path
integral is the exponential of the
cutting and pasting manifold of the complex solution to the
Einstein equation with  two
corresponding Fourier transformations
[2][4]. The two Fourier transformations are introduced for the
representation transformations of the wave function
at the equator, they are employed to make the configuration 
meaningful [2][4][6][7]. Classically, they
correspond to two Legendre transformations of the action.

For the regular compact instanton case, there are no external
imposed quantities.  Therefore, the
partition function simply counts the total number of the states,
and each state is equally probable
with probability $p_n = Z^{-1}$. Thus, the entropy is $S = -\sum
p_n\ln p_n = \ln Z$, and it is the
negative of the action [8].

For the Euclidean open Schwarzschild-type model, the boundary at
infinity is the $U(1)$ fiber bundle over the sphere with the
first
Chern number zero. In this paper we are going to perform the
calculations for its generalized versions, the
Taub-NUT and Taub-bolt-type models. In these models, the boundary
topology will be the nontrivial $U(1)$ fiber bundle over the
sphere
with a nonzero first Chern number. 

Let us begin with the Taub-NUT-anti-de Sitter metric [9][10][11]
\begin{equation}
ds^2 = b^2E (V_N(r)(d \psi + \cos \theta d \phi )^2 +4 
V_N^{-1}(r) dr^2 + (r^2 - 1) (d \theta^2 +
\sin^2 \theta d \phi^2)),
\end{equation}
where
\begin{equation}
V_N(r) = \frac{F_N(r)}{r^2 - 1} = \frac{Er^4 + (4-6E)r^2 +
(8E-8)r + 4 - 3E}{r^2 - 1},
\end{equation}
and $E$ is an arbitrary  parameter, $b^2 = -3/4\Lambda $, and
$\Lambda$ is the negative
cosmological constant. $V_N(r)$ has been chosen such that there
is a regular nut at $r= r_0 = 1$,
provided that the period of the Euclidean time coordinate $\psi$
is $\Delta \psi = 4 \pi$, $0 \leq \phi \leq 2\pi$ and
$0 \leq \theta \leq \pi$.
Then the metric is asymptotically a squashed $S^3$. The parameter
$E^{1/2}$ becomes the asymptotic
ratio between the radius of the $\psi$ direction and the radius
of the $S^2$. The action and entropy of the open regular
instanton of the
Taub-NUT-anti-de Sitter model has been calculated [11].

Since we are going to discuss the constrained instanton, the form
of $V_N(r)$ and the period conditions of the coordinates can be
considerably
relaxed. However, the motivation of our discussion is not to
exhaust  the vacuum models with the
given asymptotic topology. For simplicity, we  just restrict our
study to the case (3). We let the time identification period
$\beta$, i.e.
$\Delta \psi$, to be a free parameter. In order to construct a
closed manifold, one has to set $\Delta \psi = 2 \Delta \phi /n$,
where
$\Delta \phi$ is the period of $\phi$, then the boundary
topology is $S^3/\it Z_{\rm n}$ with the first Chern number $n$.
To
some extent, we can consider
the instanton to be built as a topologically constrained
instanton.   
If $\Delta \phi \neq 2\pi$, then there are two
conical singularities at the two poles $\theta = 0, \pi$ of the
$S^2$, in addition to the conical singularities associated with
the zeros of $V_N(r)$. There
is another $U(1)$ isometry associated with the rotation angle
$\phi$ and
the reciprocal of the temperature $\Delta \phi$, in addition to
the $U(1)$ isometry associated with
the coordinate $\psi$ and the reciprocal temperature $\Delta
\psi$.

The function $V_N(r)$ has three zeros, or horizons, $r_1, r_2$
and
$r_3$. One can construct a
compact constrained instanton using the $r$-sector connecting
$r_i$ and $r_j (0 \leq i < j \leq 3)$ of the metric (3). There is
only one degree of freedom left, that is the period $\Delta
\psi$.

The Euclidean action of the gravitational field is
\begin{equation}
I = - \frac{1}{16 \pi} \int_M (R - 2 \Lambda) - \frac{1}{8 \pi}
\oint_{\partial M}K,
\end{equation}
where $R$ is the scalar curvature of the spacetime $M$, $K$ is
the trace of the second form of
the boundary $\partial M$.

The action due to the volume is
\begin{equation}
I_v = \frac{b^2 E^2}{8 \pi} (r^3_i - r^3_j - 3(r_i - r_j)) \Delta
\psi \Delta \phi.
\end{equation}

The surface gravities for the horizons are
\begin{equation}
\kappa_l = \frac{E\prod_{k \neq l} (r_l - r_k)}{4(r_l + 1)}. \;\;
(l, k = 1,2,3)
\end{equation}
The action due to the zeros $r_l (l = i,j)$ is
\begin{equation}
I_{h,l} = \frac{b^2 E}{4\pi} (r^2_l - 1) \Delta \phi  
 2\pi ( 1 - f_l),
\end{equation}
where we assume that $\Delta \psi = f_l | \beta_l|$ and $\beta_l
\equiv 2\pi \kappa^{-1}_l$. There
is no contribution from the nut $r_0$.

The action due to the conical singularities at the two poles is
\begin{equation}
I_{c} = - \frac{b^2E}{2\pi} (r_i - r_j) \Delta \psi (2 \pi -
\Delta \phi).
\end{equation}

The total action is
\begin{equation}
I = \alpha n \Delta \psi^2 + (\gamma n + \delta) \Delta \psi,
\end{equation}
where
\begin{equation}
\alpha =  \frac{b^2 E}{16 \pi} (r_i - r_j)(E( -2 + r_i +r_j) +4),
\end{equation}
\begin{equation}
\gamma = \frac{b^2E}{4}(r_i^2 + r^2_j - 2)
\end{equation}
and
\begin{equation}
\delta = -b^2E (r_i - r_j).
\end{equation}
It is noted that the last three terms in $F_N(r)$ have no effect
in the derivation of these formulas.

In contrast with the Kerr-Newman case with the boundary topology
$S^1
\times S^2$, here the total
action does depend on the time identification period $\Delta
\psi$. Thus, the constrained
instanton is obtained by determining $\Delta \psi$ through the
stationary action condition
\begin{equation}
\frac{d I}{d \Delta \psi} = 0.
\end{equation}

From eq. (14) one can derive $\Delta \psi$ and $I$ for given
parameters $E$ and $n$ for the case that all four roots are real.
There are six possible combinations for choices
$r_i$ and $r_j$. The one with maximum action is of special
interest in quantum cosmology [1]. This instanton is the seed for
the creation of the universe. Of course, our seed will not lead
to a Lorentzian universe. A complex instanton will be the seed
for the Lorentzian universe, which can be obtained from a relaxed
form of metric (3). Then one can analytically continue the
coordinates $r$ and $\theta$
to get a Lorentzian universe. However, one has to look into the
case carefully, the stationary action condition may not be
reached for the case with  complex roots. If this is the case,
then under the constraints, no constrained gravitational
instanton is available.

One can see, if one lets the Chern number approach infinity, then
the action or the entropy is proportional to $n$, and the
reciprocal of the time
identification period $\Delta \psi$, i.e. the temperature
approaches a constant.

In general, one has $\Delta \phi \neq 2\pi$, therefore two
conical singularity strings appear in
the instanton.  The
strings connect poles of the bolts and the nut.
If one believes the first principle of
the variational calculation, instead of the field equation, one
will not feel discomfort about these
conical singularities. However, if one is really disgusted with
these
singularities, one can set $\Delta
\phi = 2\pi$ as a constraint, then there is no free parameter
left for the horizon singularities. The
entropy is no longer the negative of the action.

In a very similar way, we can perform the calculation for
the Taub-Bolt-anti-de Sitter family. Their metrics take the same
form
(3), except that $F_N(r)$ is replaced by the following $F_B(r)$
\begin{equation} 
F_B(r,s)= Er^4 + (4-6E)r^2 + [ -Es^3 + (6E - 4)s + (3E-4)s^{-1}]r
+ 4 -3E,
\end{equation}
where
\begin{equation}
E = \frac{2ns -4}{3(s^2 - 1)}
\end{equation}
and $s$ is a parameter.
There are four horizons. Eqs. (6)-(13) remain the same. The
calculation of the action and entropy
for its regular open
model has been done in [11].

If one lets $E$ approach zero while keeping $b^2E = 1$, then one
obtains the Taub-NUT or
Taub-bolt
model. For the Taub-NUT case, one has $n=1, r_i = 1, r_j = -1$
and $\beta = 2\pi, S = - I =
2\pi$. For the Taub-bolt case, one has $n = 1, r_i = 2, r_j =1/2$
and $\beta =5\pi/4, S = - I =
75\pi/128$. As expected, the results differ from those for
regular instantons using background
subtraction
method [12].

The  Taub-NUT Lorentzian universe  is the imaginary $r$-section
of  metric (3) with 
\begin{equation}
V_N(r) =\frac{4(r^2 - 2i\eta r + 1)}{(r^2 - 1)},
\end{equation}
 where $\eta$ is a parameter. The seed for its creation is the
$r$-sector
between $r_{(i, j)} = i(\eta \pm (\eta^2 + 1)^{1/2})$. No
constrained instanton exists for the
constraint with  free parameter $\Delta \psi$. One can set
$\Delta \phi = 2\pi$ or $\Delta \psi = 4 \pi$ as an extra
constraint and then obtain the instanton. Then one can calculate
the action through eqs. (6)-(13)
letting $b^2E = -1$ and $E \rightarrow 0$. The action is
complex. The creation probability is
the exponential of the negative of the real part of the action.
The action is no longer the negative of
the entropy.

By letting $n$ approach zero, and rescaling the coordinates, only
$\gamma$ term survives in
the action, and we rederive the result of the Schwarzschild-type
model [4].

The discussion can be straightforward generalized into the
Taub-NUT-Kerr-Newman-anti-de
Sitter and Taub-Bolt-Kerr-Newman-anti-de Sitter models.

\vspace*{0.3in}
\rm

\bf Acknowledgments

\vspace*{0.3in}

\rm I would like to thank J. Narlikar of IUCAA and D. Lohiya of
Delhi University for their hospitality.
 
\vspace*{0.3in}

\bf References:

\vspace*{0.1in}
\rm

1. J.B. Hartle and S.W. Hawking, \it Phys. Rev. \rm \bf D\rm
\underline{28}, 2960 (1983).

2. Z.C. Wu, \it Int. J. Mod. Phys. \rm \bf D\rm\underline{6}, 199
(1997).

3. Z.C. Wu, \it Gen. Rel. Grav. \rm \bf 30,\rm 1639 (1998), 
hep-th/9803121.
 
4. Z.C. Wu, gr-qc/9810077.

5. G.W. Gibbons and S.W. Hawking, \it Phys. Rev. \bf D\rm
\underline{52}, 2254 (1995).

6. S.W. Hawking and S.F. Ross, \it Phys. Rev. \bf D\rm
\underline{52}, 5865 (1995).

7. R.B. Mann and S.F. Ross, \it Phys. Rev. \bf D\rm
\underline{15}, 2725 (1977).

8. G.W. Gibbons and S.W. Hawking, \it Commun. Math. Phys. \rm 
\underline{66}, 291 (1979).

9. D.N. Page and C.N. Pope, \it Class. Quant. Grav.
\rm\underline{3}, 249 (1986).

10. D.N. Page and C.N. Pope, \it Class. Quant. Grav.
\rm\underline{4}, 213 (1987).

11. S.W. Hawking, C.J. Hunter and D.N. Page, hep-th/9809035.

12. S.W. Hawking and C.J. Hunter, hep-th/9808085.

\end{document}